\documentclass[conference]{IEEEtran}
\IEEEoverridecommandlockouts
\usepackage{cite}
\usepackage{amsmath,amssymb,amsfonts}
\usepackage{algorithmic}
\usepackage{graphicx}
\usepackage{multicol}
\usepackage{textcomp}
\usepackage{xcolor}
\usepackage{comment}
\usepackage{caption}
\usepackage{subcaption}
\usepackage{booktabs}
\usepackage{color,algorithm}
\usepackage{cite}  
\begin{document}

\title{Characterizing the Computational and Memory Requirements of Virtual RANs  
}

\author{\IEEEauthorblockN{Somreeta Pramanik}
\IEEEauthorblockA{
Politecnico di Torino\\
Torino, Italy \\
somreeta.pramanik@polito.it}
\and
\IEEEauthorblockN{Adlen Ksentini}
\IEEEauthorblockA{EURECOM \\
Sophia Antipolis, France\\
adlen.ksentini@eurecom.fr}
\and
\IEEEauthorblockN{Carla Fabiana Chiasserini}
\IEEEauthorblockA{Politecnico di Torino \\
Torino, Italy \\
carla.chiasserini@polito.it}
}

\maketitle

\begin{abstract}
The virtualization of radio access networks (RANs) is emerging as a key component of future wireless systems, as it brings agility to the RAN architecture and offers degrees of design freedom. In this paper, we investigate and characterize the computational and memory requirements of virtual RANs. To this end, we build a virtual RAN test-bed leveraging the srsRAN open-source  mobile communication platform and general-purpose processor-based servers. Through extensive experiments, we profile the consumption of computing and memory resources, and we assess the system performance. Further, we build regression models to predict the system behavior as the number of connected users increases, under diverse  radio transmission settings. In so doing, we develop a methodology and prediction models that can help designing and optimizing virtual RANs.  
\end{abstract}


\section{Introduction}
 Virtualization of the Radio Access Network (RAN) is well-recognized as a key technology to accommodate the ever-increasing mobile data traffic of emerging applications with stringent requirements \cite{vRAN}.  By exploiting software defined networking  and network function virtualization, the Cloud RAN (C-RAN) \cite{cloudran} concept has evolved towards the Virtual RAN (vRAN) paradigm \cite{cloudtooran}, where  a virtual eNB  centralizes a softwarized radio point of access  stack 
into a computing edge infrastructure. The majority of     the baseband processing of a virtual radio point of access (vRPA) is executed in virtual network functions running in computing platforms, typically located at edge servers close to the antennas. 
In so doing, a vRPA is able to adapt to time varying/non-uniform data traffic demands,   as new vRPAs can be added and upgraded easily, thereby improving scalability and easing network maintenance while reducing site lease costs, energy usage, and maintenance expenses. Moreover, due to the virtual network functions being implemented in edge servers, a vRPA can reduce latency, ensure highly efficient network operation and service delivery, and offer an improved user experience. 


Importantly, vRAN has also become a cornerstone technology for the realization of the emerging Open Radio Access Network (O-RAN) paradigm \cite{oRAN}. Indeed, the level of virtualization and flexibility that characterize a vRAN make it a perfect fit for the openness and intelligence concepts that are at the basis of the O-RAN architecture \cite{intelligentoran}. 
It is therefore expected that open standard radio frequency (RF) interfaces, combined with vRAN technologies, will further increase  operational savings and increase the scalability of radio access networks. 

However, the increased network flexibility and programmability allowed by vRANs come at the cost of a higher consumption of computing and memory resources at the network edge \cite{vrain}. 
 In particular, computing resources are typically pooled inefficiently since most of the implementations over dimension computing capacity to cope with peak demands in real-time workloads \cite{cares,vranbenefits}. It follows that  the gains currently  attainable by a vRAN are far from optimal, preventing its deployment at scale. 
 
 To solve the above issues, it is critical to dynamically adapt the  resources allocation  to the temporal variations of the demand across vRPAs \cite{cloudiq}. 
 Towards  this goal, 
a first, fundamental step is to gain a better  hands-on understanding of the behavior of vRPAs and the relation between radio and computing/memory resource dynamics, as well as their dependency on such factors  as radio channel conditions and user's traffic demand.  
While the studies in  \cite{cloudiq,rtopex,icc2019,perfcran1,perfcran2, cares,compoutage,fluidran,carem,vrain,mobicom21} have focused on the optimization of vRANs through experimental work or by designing analytical models and algorithmic solutions,   
to  our knowledge,  this paper is the first attempt to  characterize the computing and memory resource consumption of vRANs under diverse settings, 
as the number of connected users increases. 

In particular, we address the following research questions:

{\em 1. What are the computing and memory requirements of a vRAN, as different settings in terms of number of occupied resource blocks and type of modulation and coding scheme are adopted?}

{2. \em How do the computing and memory requirements of a vRAN change  as the number of connected users varies?}

\noindent
We answer these questions by investigating the behavior of a vRPA using a test-bed  implementation and conducting an extensive measurement campaign. In particular, we leverage an srsRAN implementation of an eNB and investigate 
its CPU and memory consumption under different experimental settings. 

Our main contributions are as follows.
\begin{itemize}

\item We develop an srsRAN-based experimental test-bed and perform extensive experiments, in order to profile the performance limits of the eNB in terms of processing, memory, and throughput. We show that the CPU utilization of the eNB increases with the modulation and coding (MCS) index, number of occupied Resource Blocks (RBs), and importantly, with the number of connected users.

\item Using empirical data, we define regression models to predict  the percentage of CPU utilization  and memory consumption of the virtual eNB, as the number of connected users varies. In so doing, we obtain a prediction accuracy of 99\% for both CPU and memory utilization.  These approximated models provide real-world insights and key inputs to formulate, design, and evaluate optimized resource management in vRANs. 
\end{itemize}

The rest of the paper is organized as follows. Section\,\ref{sec:related works} discusses the related work and highlights the novelty of our study.  Section\,\ref{sec:testbed} introduces the design and implementation of our vRAN test-bed, while Section\,\ref{sec:results} presents experimental results and our empirical models. Finally, Section\,\ref{sec:conclusion} concludes the paper and presents possible directions for future research.    

\section{Related Work}\label{sec:related works} 
Owing to the intricate relationship between radio and computing resource dynamics, and the advantages offered by vRANs, several works have aimed 
at investigating and optimizing such a virtual system. While the studies in \cite{cloudiq, rtopex, icc2019, perfcran1, perfcran2} focus on evaluating the performance of a vRAN through experiments,  the works in \cite{cares, compoutage, fluidran, carem, vrain, mobicom21} provide an insight into the theoretical framework.

\textbf{Experimental work.} 
The work in \cite{cloudiq} is one of the first experimental studies that characterize the potential savings in compute resources when exploiting the variations in the processing load across base stations. 
Interestingly,  \cite{rtopex} presents a linear model to calculate the uplink processing time for a single user in terms of the sub-carrier load, MCS index, and  number of antennas. 
The linear model is then used to develop RT-OPEX, a C-RAN scheduling algorithm. 
The impact of MCS and SNR (radio parameters) on real-time C-RAN processing (i.e., CPU) is studied in \cite{icc2019}, along with a mathematical model  for predicting the decoding time. 
The work in \cite{perfcran1} profiles instead the  performance of a C-RAN in terms of CPU and memory usage, as the iperf transmission bandwidth increases. 
In \cite{perfcran2}, the authors investigated the CPU consumption of the baseband unit (BBU) under various conditions for the C-RANs, and characterized the computational demand in terms of throughput.  
However, no such work has investigated and characterized the performance of a vRAN in terms of CPU and memory usage 
as the number of connected users increases and under diverse settings. 

\textbf{Theoretical work.} For the sake of completeness, we also mention that quite a large literature exists on analytical models and algorithmic solutions for the optimization of a vRAN. 
In particular, the works of \cite{cares,compaware15} set a theoretical basis for CPU-aware radio resource control. In \cite{cares}, the authors aim at reducing the level of variability of the computational load, which leads to eventual computational outages (where frames are not decoded in time), by jointly optimizing the selection of the MCS index and the allocation of the physical resource blocks. 
The authors in \cite{compoutage,compaware15_2}  investigate the trade-off between the consumption of data processing resources and achievable data rates, taking into account specifically the processing requirements of forward error correction (FEC) on the uplink.
The study in \cite{compoutage} proposes  a model for computational outage and shows how it relates to channel outage. 

The above discussed works rely on the same model relating computational requirements and SNR, and they neglect variations on the arrival bit-rate load. This issue is addressed instead in  \cite{wang2021computing}, which combines real-time traffic classification and CPU scheduling in a mobile edge computing  setup. However, \cite{wang2021computing} also relies on a simplistic base-band processing model and does not include an experimental validation. 
The works in \cite{compoutage,2021} propose to enable the coordination between radio and computing resources schedulers by introducing a computationally aware MCS selection policy that reduces the computational complexity requirements, at the cost of slightly decreased spectral efficiency.  An analytical framework, FuidRAN, for the vRAN design is introduced in \cite{fluidran}, which jointly selects the function split and routing policy, tailored to the available network and computing resources. 
However,  the model is provided for  one user only. In \cite{carem}, a novel reinforcement learning framework is presented, which efficiently allocates radio resources to multiple users in terms of link, MCS index, RBs and airtime for packet transmissions in heterogeneous vRANs. 
A related relevant contribution is also given in \cite{vrain}, which designs a solution, named as vrAIn, that dynamically learns the optimal allocation of computing and radio resources in order to meet the target level of quality of service.    
A novel pipeline architecture for 5G distributed units (DUs) is then presented in \cite{mobicom21} to guarantee a minimum set of signals that preserve synchronization between the DU and its users, during computing capacity shortages. This study relies on techniques that require predictable computing to provide carrier-grade reliability.

In conclusion, there is no such work which characterizes the computing requirements of vRANs with respect to contextual dynamics (for instance, traffic load and number of users).  
Importantly,  designing resource allocation schemes without knowing such requirements may severely hamper the system performance in terms of throughput.

\begin{figure*}[tb]
\centering
\subfloat[]
{\includegraphics[width=0.58\textwidth]{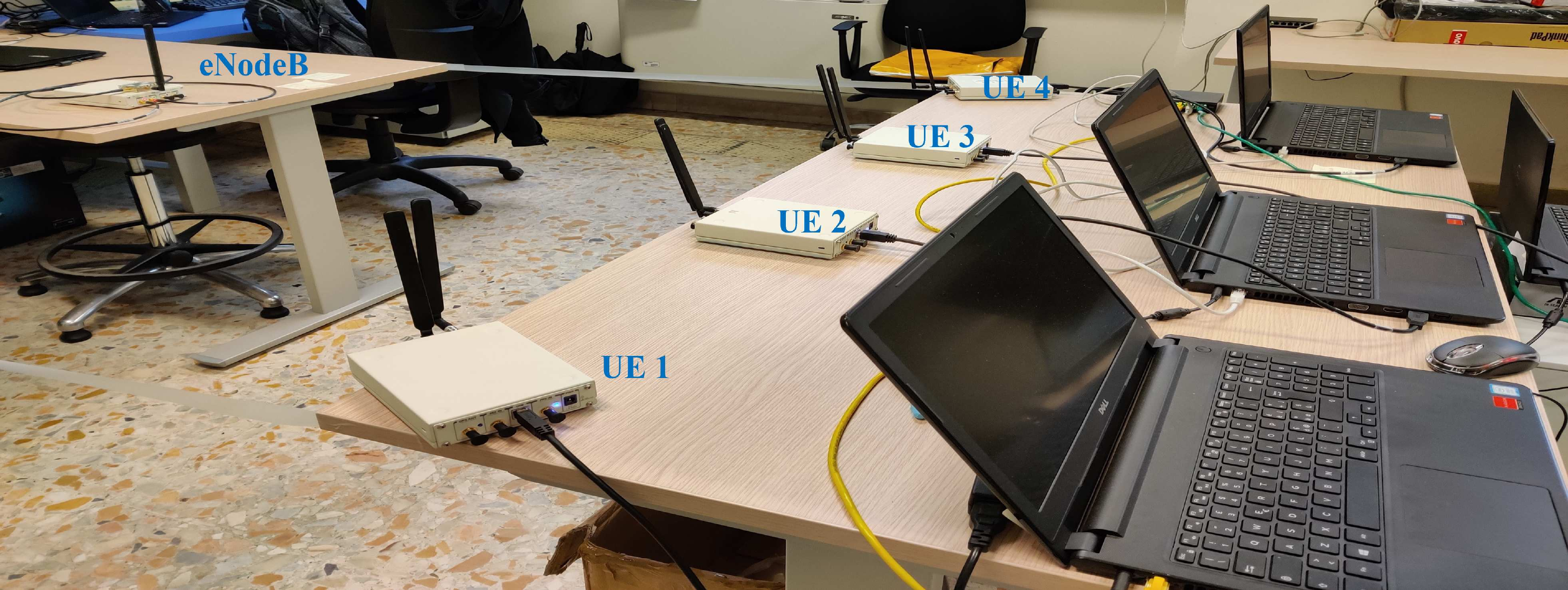}}\hfil
\subfloat[]
{\includegraphics[width=0.35\textwidth]{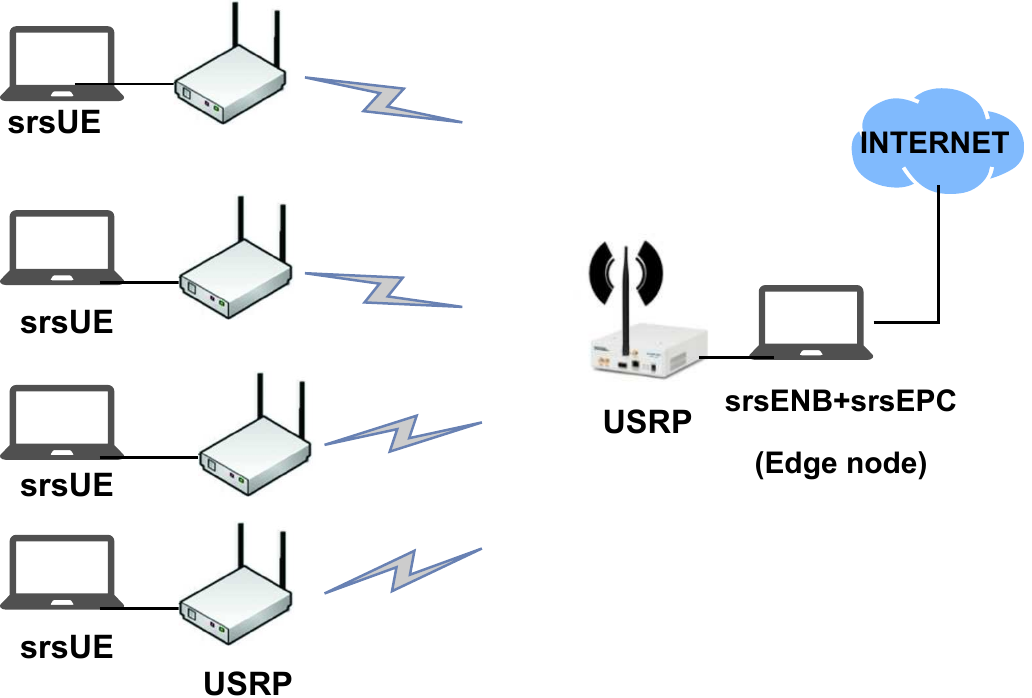}}
\caption{vRAN test-bed implementation using srsRAN: (a) snapshot of our test-bed highlighting the edge node hosting the virtual eNB and EPC, and four UEs; (b) test-bed architecture including UEs, virtual eNB and virtual EPC.}
\label{test-bed}
\end{figure*}

\section{vRAN Test-bed}\label{sec:testbed}
We  now introduce our vRAN test-bed using srsRAN, detailing the test-bed architecture and configuration, and the adopted experimental methods. 

\subsection{Test-bed architecture}\label{subec:testbed-arch}
Figure\,\ref{test-bed}(a) provides a snapshot of the test-bed we developed, while Figure\,\ref{test-bed}(b) represents its architecture. We leverage software defined radio (SDR) interfaces enabling point-to-point communications between vRPA and User Equipment (UE). An vRPA implements the necessary processing stack to transfer data to/from UEs. In our case, the vRPA acts as a virtual eNB implemented at the edge of the network. The connectivity between the vRPA and UEs is supported by means of an LTE radio link implemented using the srsRAN \cite{srslte} -- an open-source SDR LTE stack implementation offering Evolved Packet Core (EPC), eNB, and UE applications. It is compliant with LTE Release 9 and supports up to 20-MHz bandwidth channels as well as transmission modes from 1 to 4, all using the frequency division duplexing (FDD) configuration. 

As RF front-end, Ettus Universal Software Radio Peripheral (USRP) B210 devices are used to perform up/down-conversion, filtering, amplification and AD/DA conversion of the UE and eNB LTE signals. All the RF front-ends are connected to the vRPA and the UEs via  USB. 
Then, physical layer is implemented through a set of OFDMA-modulated channels, using  RB filling across ten 1-ms subframes forming a frame. RBs assigned to UEs by the MAC layer are modulated and encoded with an MCS that depends on the user’s Channel Quality Indicator (CQI), a measure of SNR that is locally available at the UEs for uplink transmission and is reported periodically by UEs.  
The modulation order can vary up to 256-QAM, while FEC is employed using LDPC or Turbo codes with coding rate of 1/2, 2/3, or 3/4 \cite{3gpp}. At the MAC layer, an automatic repeat request error control is in place, i.e., an unsuccessfully transmitted packet can be resent till a maximum number of allowed re-transmission attempts.

The edge host and the mobile terminals are each installed in Ubuntu 18.04 systems. The edge host is equipped with an Intel i7-7700HQ 4-cores CPU and 8\,GB of DDR4 RAM, while the UEs  feature an Intel i7-8550U 4-cores CPU and 16\,GB of DDR4 RAM. Each Ubuntu system is connected to USRP B210 boards using USRP Hardware Driver v3.15. 
In order to faciltate the experiments, all performance management features in the BIOS (e.g., Intel@TurboBoost, Hyper-thread control, Intel SpeedStep) are enabled and C-states have been turned off.  
Moreover, the real-time thread priorities are enabled in the srsRAN as the applications (srsENB and srsUE) are executed with root privileges. A set of threads are created in srsRAN for performance and priority management reasons. The threading architecture of the physical layer implementation is motivated by the stringent latency requirements. 
Also, we monitor the level of CPU consumption and ensure that, during our experiments, the allocated CPU is sufficient to keep up with the required data rate so as to avoid severe system failures during the radio data transfer. 
Finally, in order to establish a stable connection, we set the  transmit gain (tx\textunderscore gain) at the eNB to its maximum value. 

\subsection{Monitoring the srsRAN eNB and  UEs}
To monitor the behavior and track the performance of the vRAN entities, we leverage some of the useful features of srsRAN (e.g., detailed log system with per-layer log levels, MAC layer Wireshark packet capture, command-line trace metrics, detailed input configuration file). A configuration file is provided to set parameters such as downlink carrier frequency and log or packet capture options, making the software easy to use. Moreover, we have relied on a useful feature of srsUE which provides real-time traces.

The srsENB application metrics are generated once per second by default. Metrics are provided on a per-UE basis for the downlink  and uplink, respectively. 	
The eNB is configured in band 7 (FDD) and the transmission bandwidth has been set to $10$\,MHz, 
corresponding to $50$ RBs.
In order to determine the successful connection between eNB and UE, the RRC states are observed. Specifically, when the UEs are successfully paired to the eNB, the RRC connection setup message is seen. We have also saved the UE and eNB logs to verify such entities' status.

As experimental set-up,  we connected 30\,dB attenuators to the antennas of each network node; furthermore, the UEs were placed close enough to the eNB so as to ensure high values of SINR ($\geq 25$\,dB). We focus on downlink data transfer and used iperf for data packet  generation. 

\section{Experimental Results and Empirical Models}\label{sec:results}
We present here the performance of the vRPA, i.e., the srsRAN eNB, in terms of CPU and memory utilization as the number of occupied RBs and the MCS index vary,  when a single UE or multiple UEs are connected. The results have been obtained by averaging over 10 experiments; in every plot, both the average value of the presented performance metric and the corresponding 95\% confidence  interval are shown.

\subsection{CPU utilization}

The CPU usage is analysed using an Intel Core i7-7700HQ 2.80\,GHz CPU. The CPU utilization is calculated using the {\tt top} process in Linux, which is widely used and provides a dynamic real-time view of a running system managed by the kernel. Specifically, the percentage of  consumed CPU is collected by sampling the {\tt top} reports every second, over a period of 200 seconds. 

\begin{figure}[htbp]
\centerline{\includegraphics[width=0.4\textwidth]{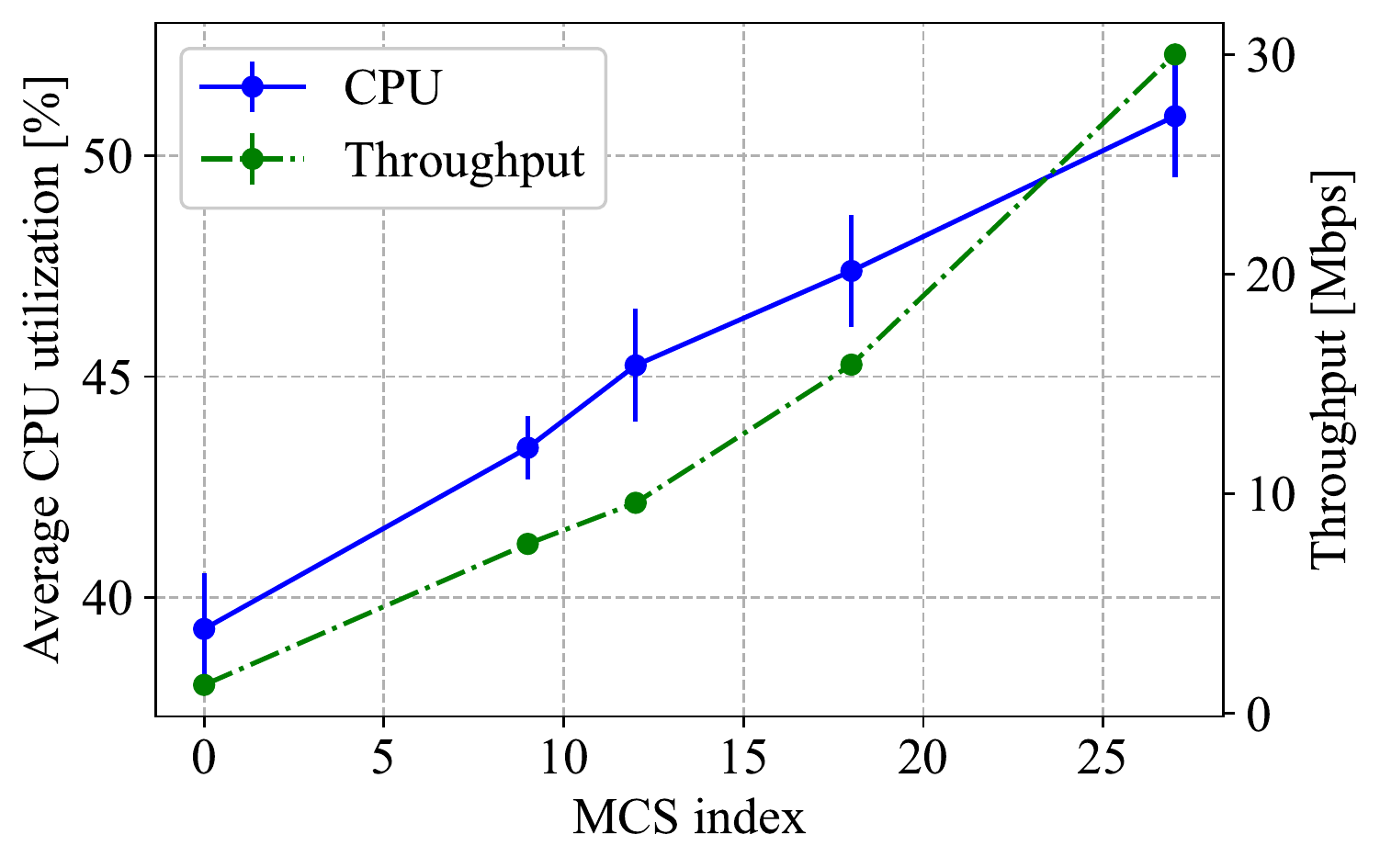}}
\caption{CPU utilization and UDP downlink throughput of the virtual eNB,  for different values of the MCS index and a single connected UE.}
\label{MCS}
\end{figure}

Figure\,\ref{MCS}  shows the average CPU utilization (left-hand side y-axis) and the throughput (right-hand side y-axis) obtained over 10 iterations of the virtual eNB, as the MCS varies and for a single connected user. 
For this experiment, UDP traffic is generated at  the eNB at $30$ Mbps and sent to the UE, setting the number of allocated RBs to $50$. Also, we set  the  transmission gain to its maximum value, thus ensuring that the SNR does not drop below $32$\,dB.  From the plot, we can observe that, as also shown in \cite{vrain}, the CPU utilization of the virtual eNB increases as the MCS index grows from $0$ to $27$.  
Further, the consumption of computing resources, which is mainly due to the modulation, demodulation, coding and decoding operations, is quite significant in absolute terms: as an example, for MCS$=27$, a single user consumes around $51$\% of a single CPU of the edge node. 
Using empirical data, we found that the CPU utilization of the virtual eNB can be well approximated as a linear increasing function of the MCS, i.e., 
$\mbox{CPU[\%]}=0.429 \times \mbox{MCS}+39.58$. 

\begin{figure}[htbp]
\centerline{\includegraphics[width=0.4\textwidth]{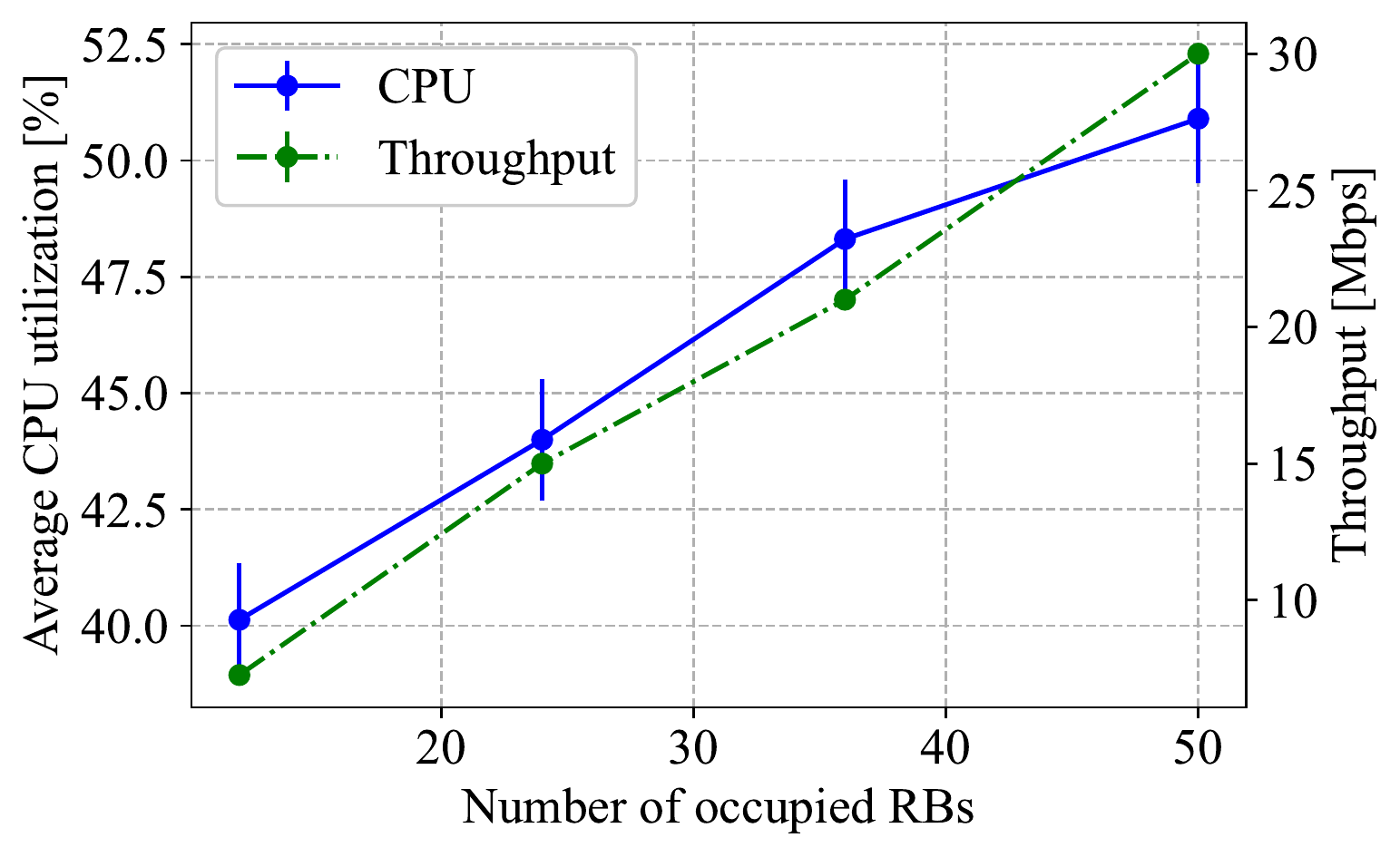}}
\caption{CPU utilization and UDP downlink throughput of the virtual eNB vs the number of occupied RBs, for a single connected UE and MCS=$27$.}
\label{RBS}
\end{figure}

Figure\,\ref{RBS}  shows the CPU utilization (left-hand side y-axis) of the virtual eNB as the number of allocated RBs varies from $12$ to $50$, for a single user and MCS$=27$. 
The downlink traffic load is set to $9$ (for $12$ RBs), $15$ (for $24$ RBs), $21$ (for $36$ RBs) and $30$ (for $50$ RBs) Mbps, respectively. We notice that the CPU utilization increases as the number of occupied RBs increases, with a maximum of $51$\% for a single UE. 
A higher number of occupied RBs leads to the user transmitting at a higher rate, which results in a higher computational resource consumption.  
From the experimental data, we found that the CPU utilization of the eNB can be well approximated as a linear increasing function of the number of occupied RBs, i.e.,
    $\mbox{CPU[\%]}=0.2892 \times \mbox{No\_RB}+37.02$. 
We remark that the provided  approximation functions can  help in interpolating the average CPU utilization with different radio configurations. 

\begin{figure}
\centerline{\includegraphics[width=0.4\textwidth]{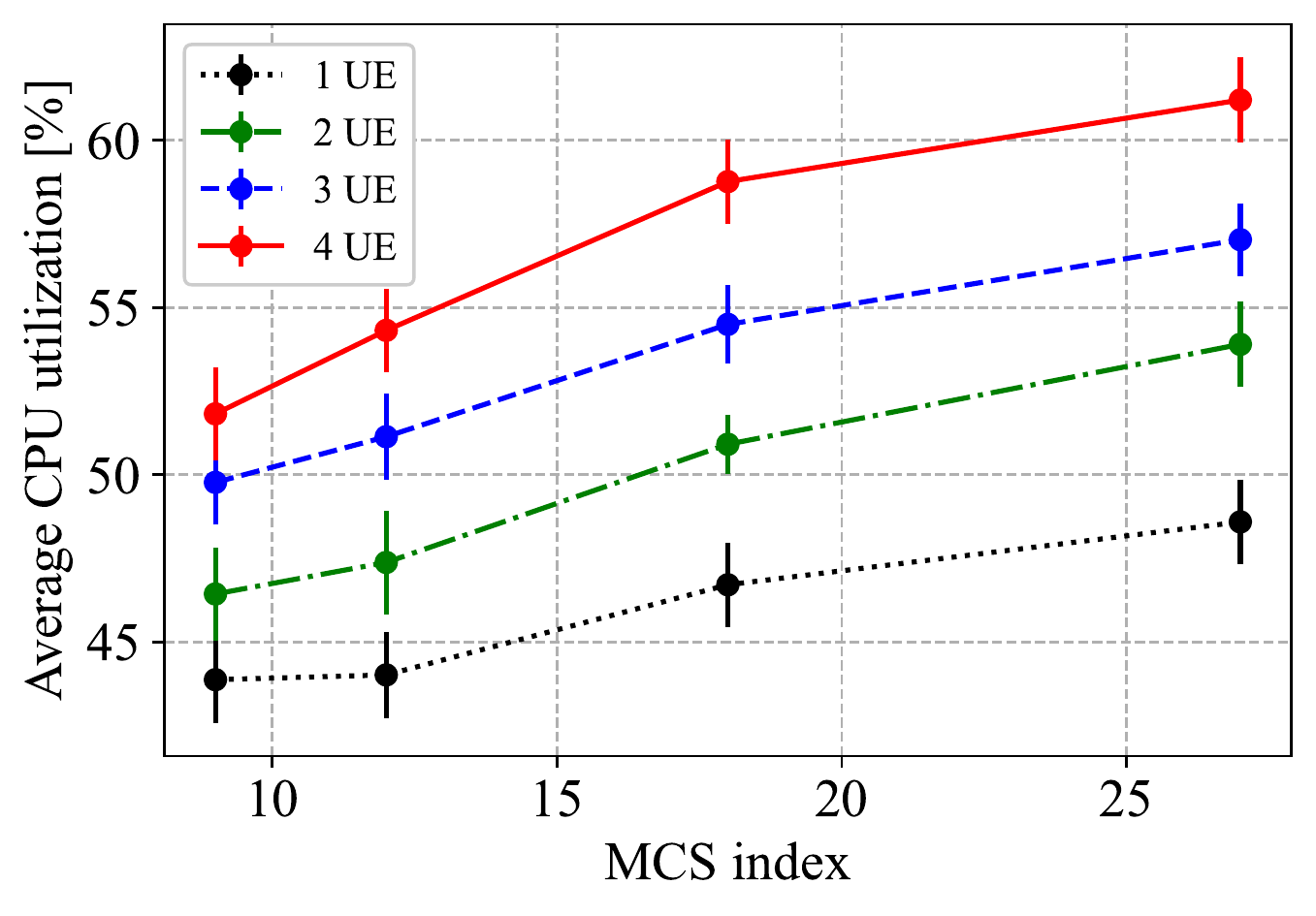}}
\caption{CPU utilization of the virtual eNB vs the MCS index, for a varying number of connected UEs and a total number of occupied RBs  equal to $36$.}
\label{fig:4user_mcs}
\end{figure}

We are now interested in how the computing resource consumption varies as the number of users connected to the eNB changes. It is indeed a fact that the number of served UEs is rapidly increasing, and that cellular networks will have to support a massive number of users. 
Figure\,\ref{fig:4user_mcs} presents the CPU utilization (left-hand side y-axis) of the virtual eNB as  the MCS index varies, for different numbers of  users. For this experiment, the overall maximum number of RBs that can be used is set to $36$, downlink traffic is generated at $21$\,Mbps, and the tx\textunderscore gain is set to its maximum value, so that the SNR is always above $28$\,\,dB for all the UEs. 
In this scenario, an interesting behavior emerges:  for a fixed value of the MCS index,  the average CPU consumption of the eNB increases significantly as the number of users increases, although the traffic load is kept constant. As an example, for MCS$=27$, the average CPU consumption with four UEs is $62$\% of a single CPU, i.e., about 30\% more than with one UE.

\begin{figure}
\centerline{\includegraphics[width=0.4\textwidth]{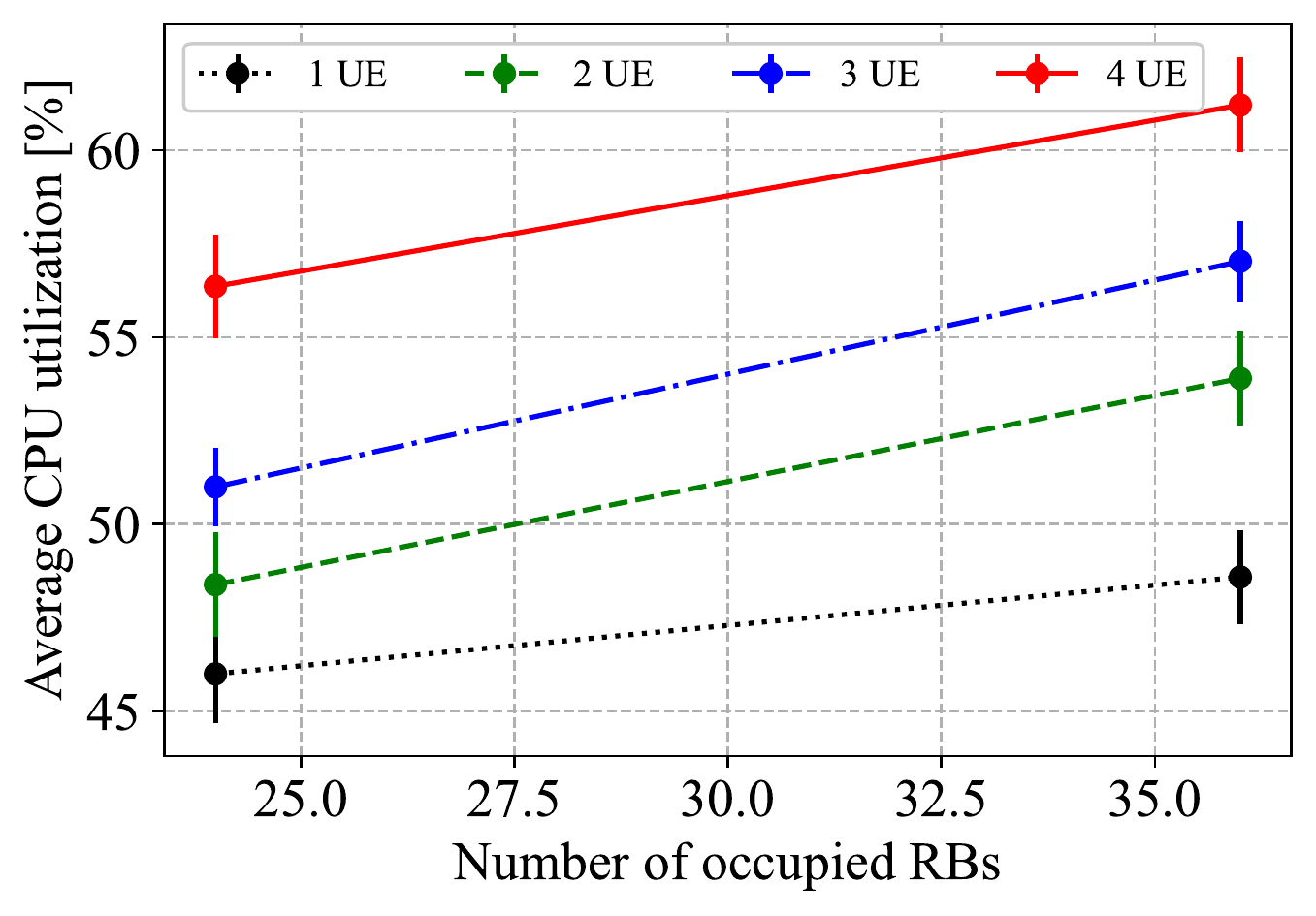}}
\caption{CPU utilization of the virtual eNB vs number of occupied RBs, for a varying  number of connected UEs and  MCS$=27$.}
\label{fig:4user_rbs}
\end{figure}

\begin{figure}
\centerline{\includegraphics[width=0.4\textwidth]{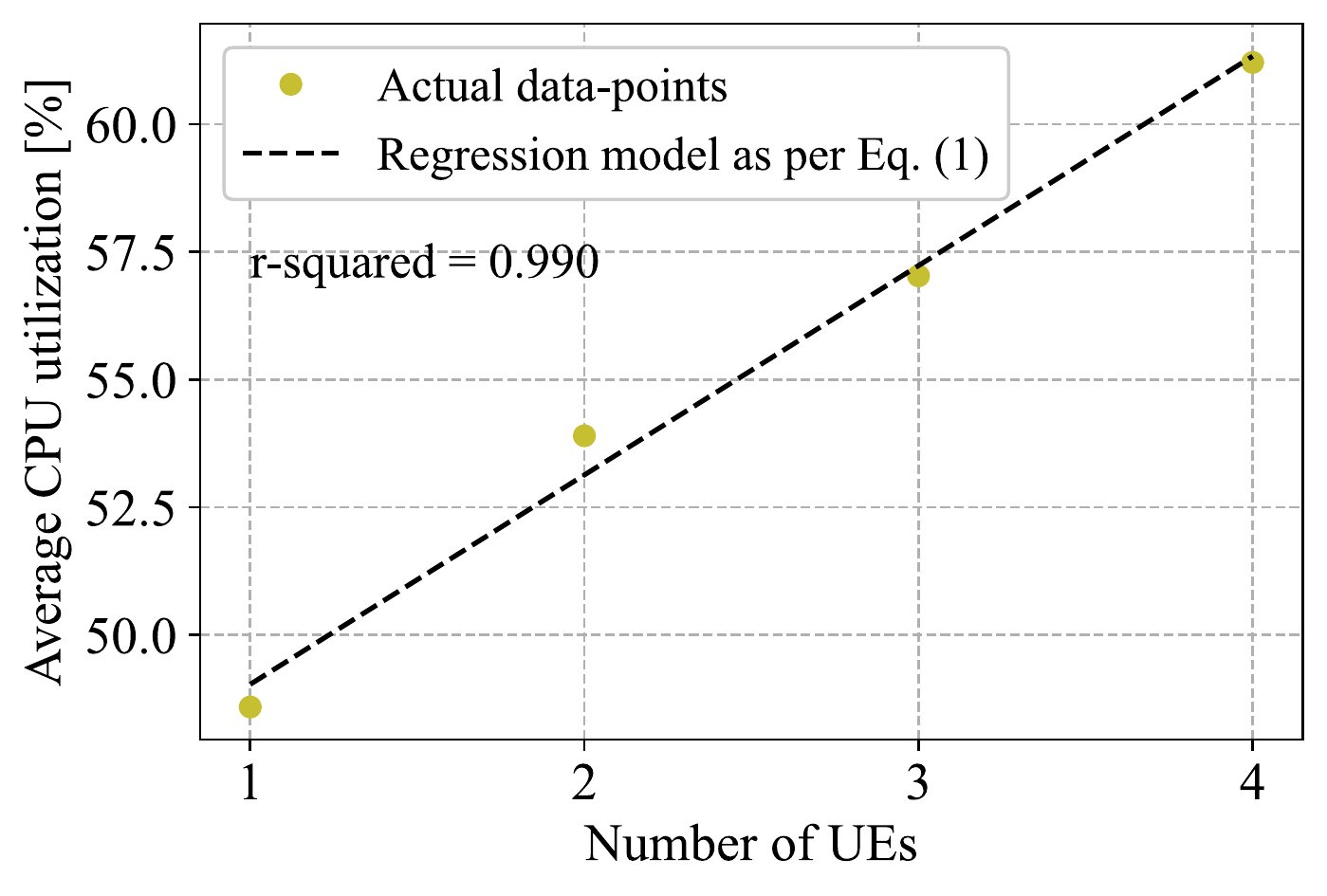}}
\caption{Regression plot as the number of UEs varies, for $36$ occupied RBs and MCS$=27$.}
\label{fig:reg_ues}
\end{figure}

\begin{figure}
 \centering
 \begin{subfigure}[b]{0.5\textwidth}
   \centering
   \includegraphics[width=0.65\linewidth]{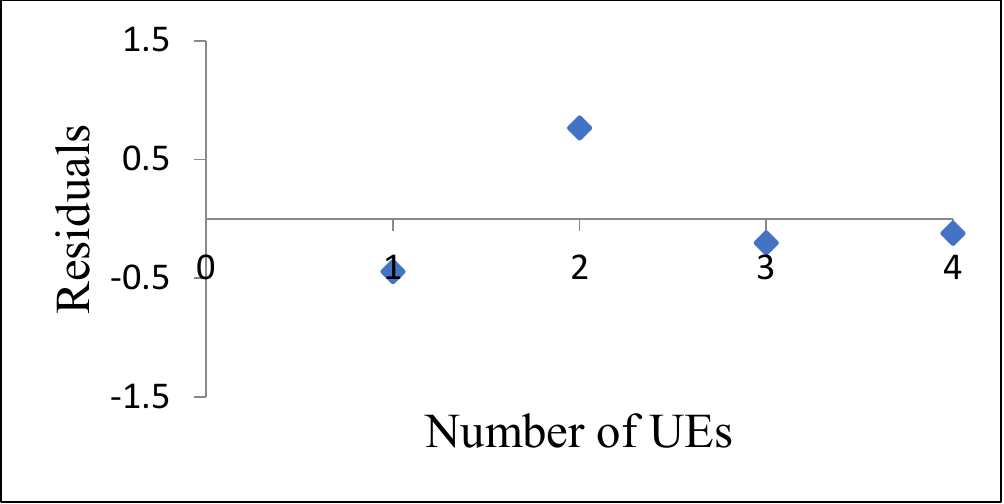}
   \caption{}
   \label{fig:res_4ues_a}
 \end{subfigure}%
 \hfill
 \begin{subfigure}[b]{0.5\textwidth}
   \centering
   \includegraphics[width=0.65\linewidth]{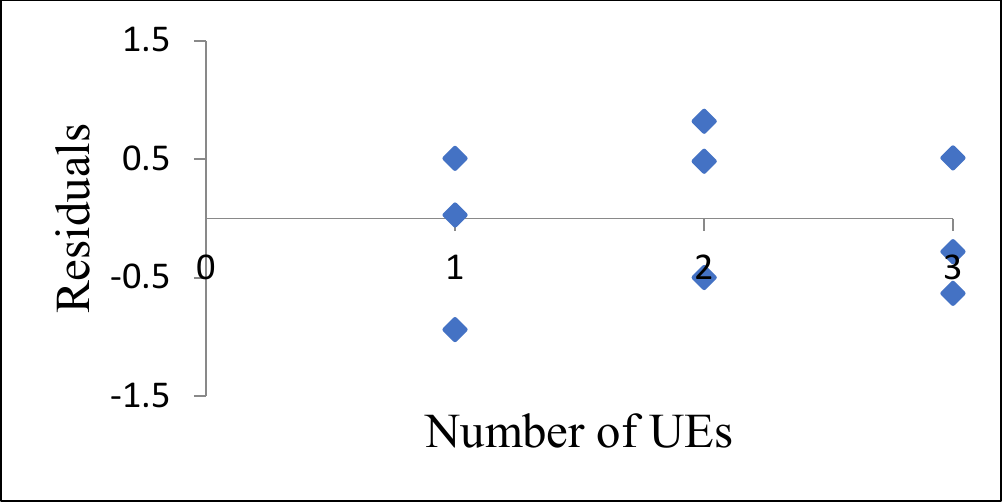}
   \caption{}
   \label{fig:res_4ues_b}
 \end{subfigure}
   \caption{(a) Residual plot for the regression model in Eq.\,(\ref{eq:cpu-n}), with 36 occupied RBs and  MCS$=27$; (b) Residual plot for the regression model in Eq.\,(\ref{eq:cpu-n-m}), for different values of the MCS index and $36$ occupied RBs.}
    \label{fig:res_4ues}
\end{figure}

In addition, Figure\,\ref{fig:4user_rbs} shows the CPU consumption of the virtual eNB as the number of allocated RBs varies from $24$ to $36$, for a different number of connected UEs, MCS$=27$, and maximum tx\textunderscore gain. 
The downlink traffic load for  $24$\,RBs is set to $15$\,Mbps,  while it is  $21$\,Mbps for $36$\,RBs, so that all the allocated RBs are always occupied. Interestingly, as the number of connected users grows, the CPU consumption of the virtual eNB increases linearly. 

Since the MCS index and the number of occupied RBs are always finite values that vary over a very specific range,  we are mainly interested in understanding the CPU requirements of the virtual eNB as the number of users increases.  Then, using the empirical data, we build a regression model that predicts the CPU utilization of the eNB  as the number, $n$, of connected UEs varies. For 36 occupied RBs and MCS$=27$, we obtain:  
\begin{equation}\label{eq:cpu-n}
CPU [\%]= 4.099 \times n+44.935\,.
\end{equation} 
We remark that similar  models can be built for different values of MCS index and number of occupied RBs.

Figure\,\ref{fig:reg_ues} shows the CPU utilization under the above settings, along with the  curve obtained using the regression model in (\ref{eq:cpu-n}). 
The Significance F for our model is 0.004,  which, being well below 0.05,  shows that the model can predict  correctly the behavior under study. Further, the regression output (R-squared) indicates that 99\% of the variation in CPU consumption is due to the number of UEs. Specifically, every additional UE is expected to entail about 4.1\% of increase in CPU usage at the eNB; it follows that 15 users will easily consume up to 100\% of a CPU. 

Again for 36 occupied RBs and MCS$=27$, Figure\,\ref{fig:res_4ues_a} plots the residuals, i.e., the difference in percentage between the actual value of CPU utilization and the one predicted by the regression model, obtained when the number of  UEs varies between 0 and 4.  We observe that such  residuals are always  within -0.5\% to 1\% of CPU usage, which confirms the very good accuracy of the model. 

Next, it is important to show that a linear regression model (as in (\ref{eq:cpu-n})), obtained from  experimental data,  correctly characterizes the computing requirements of a vRAN as the number of users increases.
To this end, we use all the  experimental data  obtained for a number of UEs up to three (i.e., for varying number of occupied RBs and  MCS indices),
and we predict the CPU consumption when four UEs are connected. The corresponding regression model is given by:
\begin{equation}\label{eq:cpu-n-m}
CPU[\%]= 3.9 \times n+0.369 \times m+35.658
\end{equation}
where $n$ is the number of connected UEs and $m$ is the adopted MCS index. A similar model can be obtained considering as explanatory variables   the number of connected UEs and the number of occupied RBs.

Tab.\,\ref{tab:table1} and Tab.\,\ref{tab:table2} report the actual CPU utilization when four UEs are connected, and the corresponding value predicted through (\ref{eq:cpu-n-m}). 
Furthermore, Figure\,\ref{fig:res_4ues_b} plots the residuals obtained for the model in (\ref{eq:cpu-n-m}):  again, all residual values  are within -1\% and 1\%. These results, along  with an F-statistic value of 0.0023,  indicates that the linear regression model well describes the behavior of CPU utilization. Similar results have been obtained for different values of the number of occupied RBs.

\begin{table}[h!]
\centering
\caption{Actual vs predicted CPU usage with 4 UEs, a varying number of occupied RBs, and MCS$=27$}
\begin{tabular}{ |c|c|c| }
 \hline
 No.\,occupied RBs & Actual CPU [\%]& Predicted  CPU [\%] \\ [0.5ex] 
 \hline\hline
 24 & 56.36 & 55.51   \\
 \hline
 36 & 61  & 60.759  \\ 
 \hline
\end{tabular}
\label{tab:table1}
\end{table}
\begin{table}[h!]
 \centering
 \caption{Actual vs predicted CPU usage with 4 UEs, different MCS indices, and 36 occupied RBs}
\begin{tabular}{ |c|c|c| } 
 \hline
 MCS index & Actual CPU [\%] & Predicted CPU [\%]\\ [0.5ex]
 \hline\hline
 12 & 54.31 &  55.55\\ 
 \hline
 18 & 58.76  & 57.71 \\
 \hline
 27 &  61 & 61.19\\
 \hline
\end{tabular}
\label{tab:table2}
\end{table}
\begin{figure}
\centerline{\includegraphics[width=0.4\textwidth]{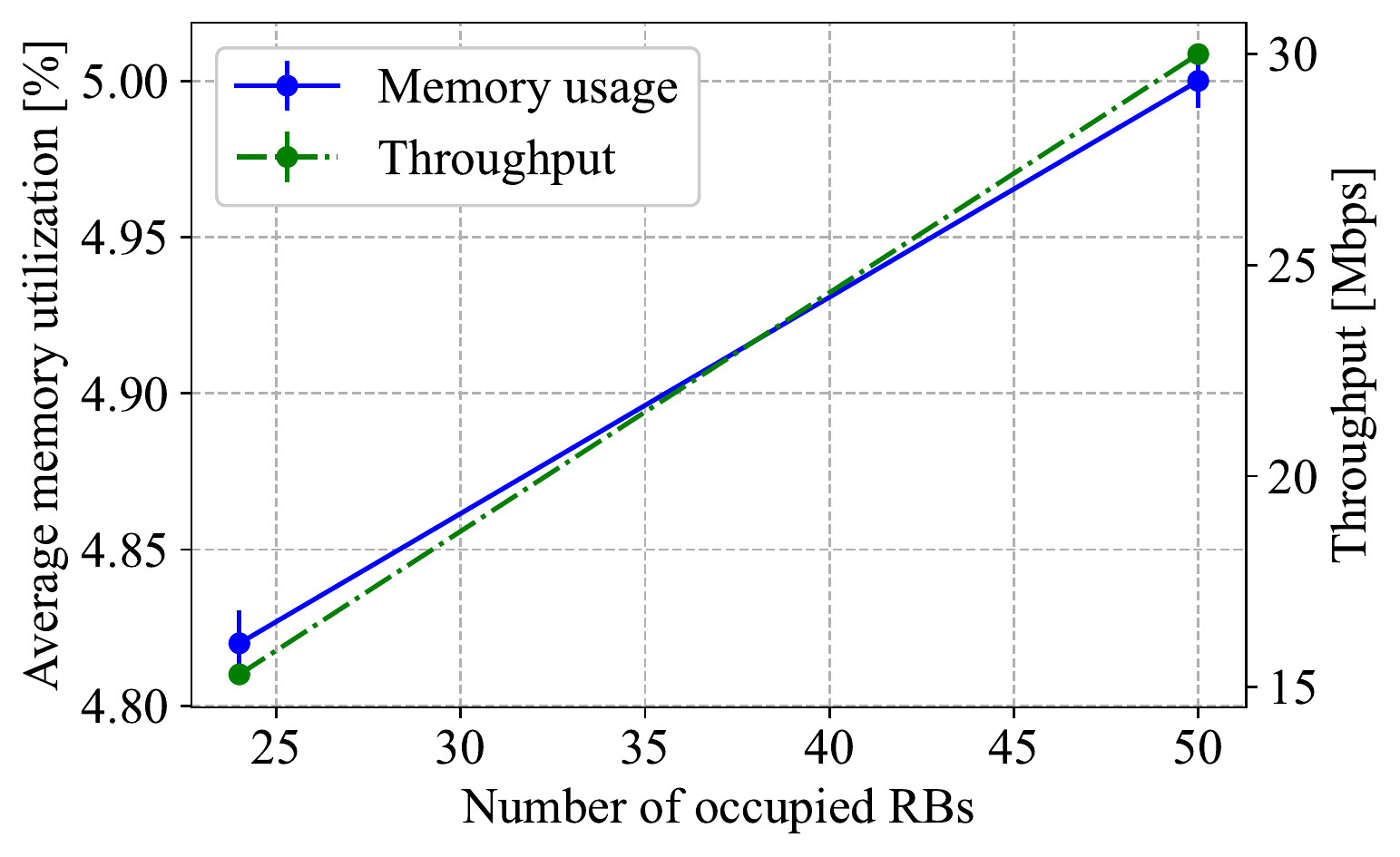}}
\caption{Memory utilization of the virtual eNB as the number of occupied RBs varies, for MCS$=27$ and one connected UE.}
\label{MEMORY_prb}
\end{figure}

\begin{figure}
\centerline{\includegraphics[width=0.4\textwidth]{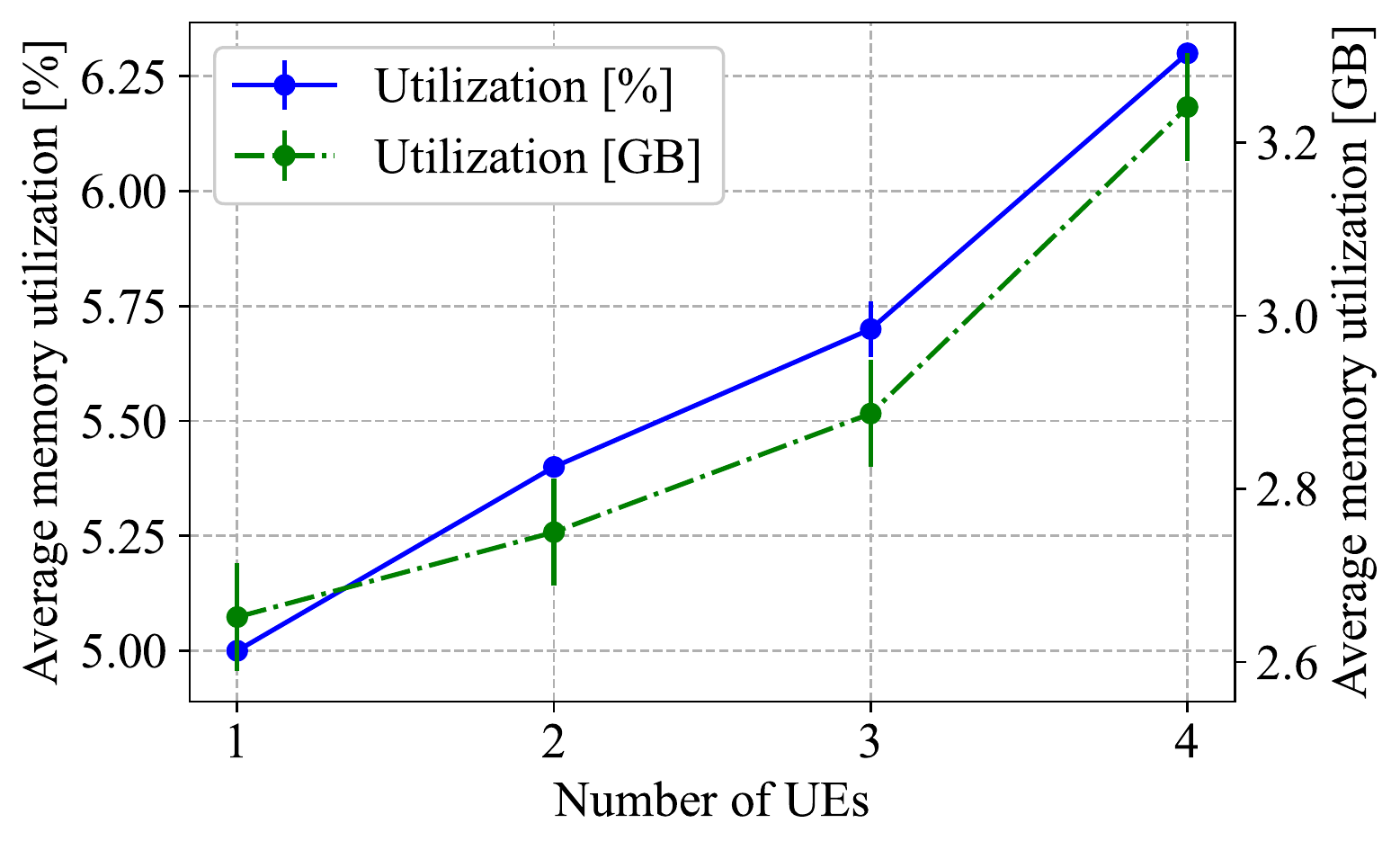}}
\caption{Memory utilization of the virtual eNB vs number of UEs, for MCS$=27$ and $50$ occupied RBs.}
\label{MEMORY}
\end{figure}

\subsection{Memory utilization} 
Beside CPU, also memory is a precious resource at the network edge. It is therefore important to investigate the memory usage of the virtual eNB for different radio configurations. 

The memory usage is monitored using the {\tt top} process in Linux, which reports both the amount of memory used by the process in GB as well as the percentage value.  As done for the CPU utilization, we profile the memory usage for different values of  MCS index, number of occupied RBs, and number of connected UEs. 
In our experiments, the downlink traffic load is  set to $30$\,Mbps  in order to ensure that all allocated RBs are actually occupied. 

Figure\,\ref{MEMORY_prb} shows the memory usage (left-hand side y-axis) as the number of occupied RBs varies from $24$ to $50$, for MCS$=27$ and only one connected UE. 
As observed for the CPU utilization, the memory usage also increases linearly with the number of occupied RBs. We also find a similar behavior as the number of UEs grows, as shown in Figure\,\ref{MEMORY} for MCS$=27$ and 50 occupied RBs. In particular, the percentage of memory utilization  (left-hand side y-axis) increases almost linearly  with the number of users, in spite of the constant traffic load. 
Furthermore, we observe that $3.24$\,GB of memory are consumed with four connected UEs, which is over 20\% more than when only one UE is connected.  

Based on the experimental data we obtained, the regression model predicting the percentage of memory utilization, for 50 occupied RBs and MCS$=27$, is as follows:
\begin{equation}
\mbox{\em Memory}[\%]= 0.42 \times n+4.55,
\end{equation}
where $n$ is the number of UEs. 
A similar linear regression model can be built for different values of number of occupied RBs and MCS index. 

\begin{figure}
 \centering
 \begin{subfigure}{0.5\textwidth}
   \centering
   \includegraphics[width=0.7\linewidth]{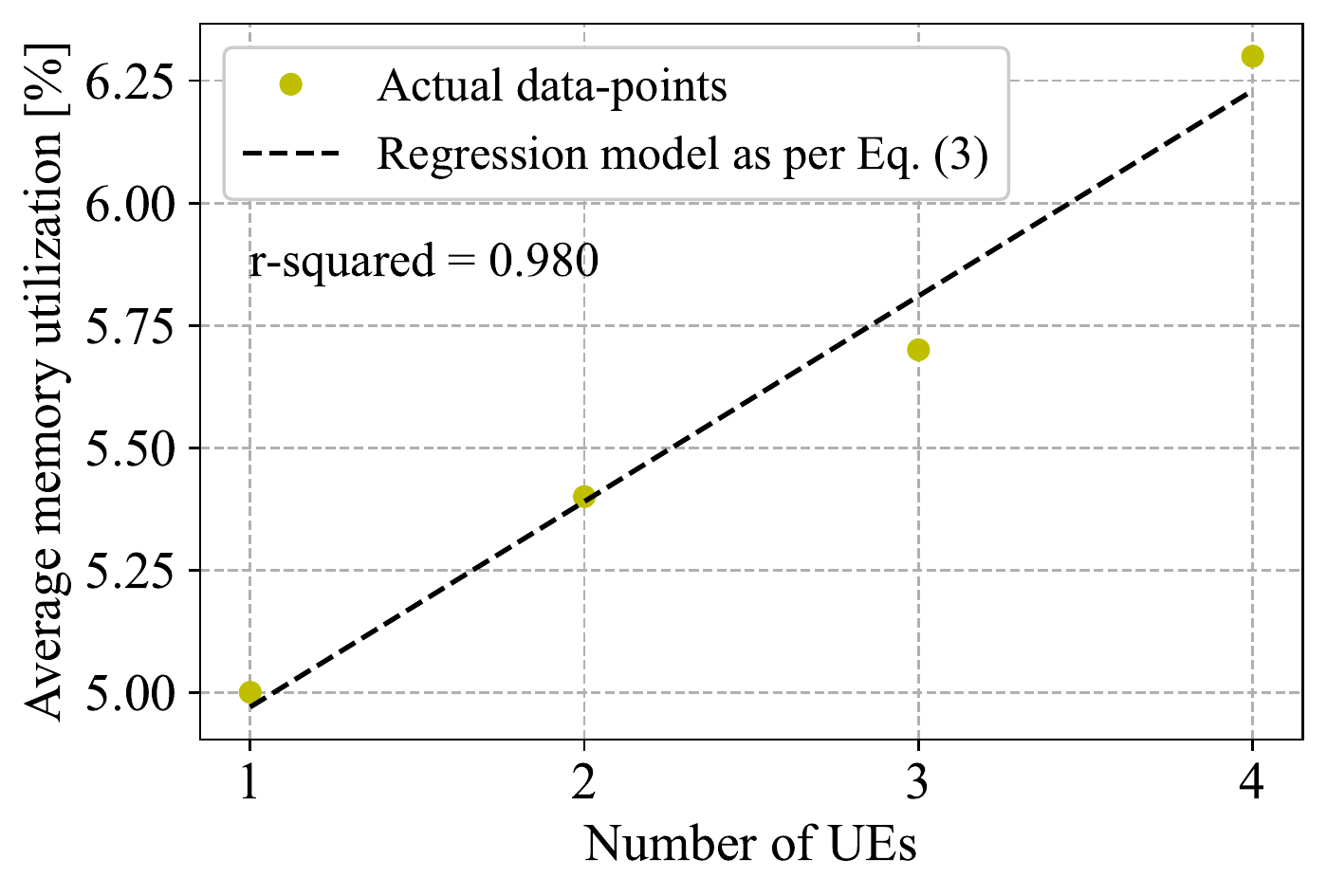}
  \caption{}
   \label{LRP3}
 \end{subfigure}
 \begin{subfigure}{0.5\textwidth}
   \centering
   \includegraphics[width=0.65\linewidth]{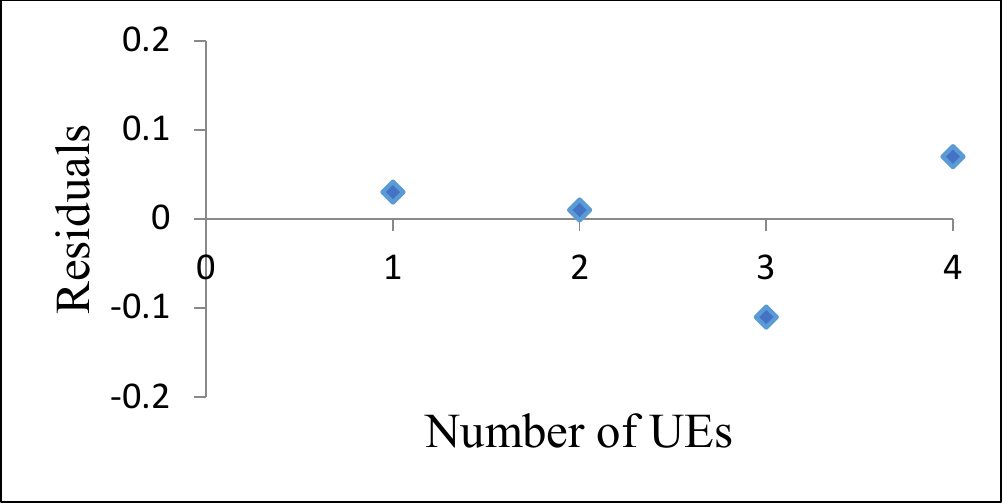}
 \caption{}
   \label{LRP4}
 \end{subfigure}
   \caption{Percentage of memory usage: (a) linear regression plot, (b)  residual plot for MCS$=27$ and $50$ occupied RBs.}
    \label{LRP_plots}
\end{figure}

The accuracy of the model is shown in Figures\,\ref{LRP3}--\ref{LRP4}, for MCS$=27$ and 50 occupied RBs. The two plots depict, respectively, the difference between the actual and predicted percentage of memory utilization, and the residuals, as the number of UEs varies. Both figures highlight that also the regression model we built for the percentage of memory utilization closely mimics the resource consumption of a vRAN.  

\section{Conclusions}\label{sec:conclusion}
Virtualized radio access networks (vRANs) are the basis of future  base stations design. To provide real-world insights and key inputs to formulate, design, and evaluate optimized resource-management problems in vRANs, we investigated and characterized the computational requirements of vRANs by developing an  srsRAN-based test-bed.  Through extensive experiments, we profiled the   CPU and memory utilization of the vRAN. Our results shed light on the vRAN behavior  across different scenarios, showing that, remarkably, the CPU and the memory utilization of the eNB increases substantially with the number of users. It is worth underlining that the result has been obtained under a constant value of traffic load and number of occupied resource blocks. Based on these empirical results, we also built linear regression models for the prediction of CPU and memory utilization as the number of users varies. To the best of our knowledge, this is the first work that thoroughly studies the computational and memory requirements of a vRAN.

We believe that our study and the obtained models can provide researchers and practitioners with real-world insights and the necessary tools for designing advanced efficient resource-provisioning and allocation strategies in vRAN systems.
As future work, we intend to exploit the results of this work to develop algorithms for the efficient and effective scaling of the vRAN functions, and the optimal settings of the vRAN parameters in the case of shortage of CPU and memory resources at the network edge.

\bibliographystyle{ieeetr}
\bibliography{ref}

\end{document}